# QUANTUM FLUCTUATIONS, THE CASIMIR EFFECT

# AND

# THE HISTORICAL BURDEN

**by**

**Gerald E. Marsh**

**Argonne National Laboratory (Ret)**
gemarsh@uchicago.edu

## ABSTRACT

It has been argued since 1948, when it was experimentally demonstrated, that the Casimir effect — where two non-charged conducting plates have a weak but measurable force on each other dependent on the inverse fourth power of the distance between them — shows the reality of vacuum zero-point fluctuations. This "proof" of the reality of vacuum fluctuations has been repeated in many quantum field theory books and papers subsequent to 1948. The attractive force is generally ascribed to the difference in zero-point energy of the electromagnetic field between the plates and the vacuum external to them. As is well known, zero-point vacuum fluctuations are incompatible with relativistic physics and are at the root of the "cosmological constant" problem. Most texts on quantum mechanics and quantum field theory eliminate the vacuum energy by normal ordering or some other mechanism. These issues are explored in this paper and it is pointed out that a means to resolve them already exists.

# Introduction

There is some ambiguity about the term "vacuum fluctuations" and "zero-point energy" in the literature. If one is discussing the lowest energy or ground state of some quantum mechanical system, the uncertainty principle tells us that the Hamiltonian must contain the term $\hbar\omega/2$.

If an electric, magnetic or vector potential field is present in the vacuum, the vacuum expectation of its field operator will vanish, but the expectation of the square of the field operators will not, which implies there are what are often called vacuum fluctuations of the field. In quantum field theory, each point in space has this zero-point energy associated with it, thus leading to infinite energy in any finite volume.

Both Julian Schwinger and Wolfgang Pauli cast doubt on the reality of vacuum fluctuations. In Schwinger's source theory, the vacuum is "the state of zero energy, zero momentum, zero angular momentum, zero charge, zero whatever," and Pauli who stated that "it is quite impossible to decide whether the field fluctuations are already present in empty space or only created by the test bodies" and as late as 1946, he is quoted as saying that "zero-point energy has no physical reality."

It is often said that even the vacuum empty of all fields still retains the zero-point energy, whose average energy vanishes. What is left are vacuum fluctuations of the so-called virtual particles that satisfy $\Delta E \Delta t \gtrsim \hbar$ so that energy can be taken from the vacuum to allow particles to appear for very short times. These are the type of vacuum fluctuations that apply as well to the Unruh and Hawking effects, whose reality Schwinger and Pauli doubted.

However, it has been argued since 1948, when it was experimentally demonstrated, that the Casimir effect[1] — where two uncharged parallel conducting plates have a weak but measurable force on each other dependent on the inverse fourth power of the distance between them — shows the reality of vacuum zero-point fluctuations. Casimir calculated and interpreted the attractive force between these plates as being due to the quantum electromagnetic zero-point energy of the normal modes between the plates. This "proof" of the reality of vacuum fluctuations has been repeated in many quantum field theory books and papers subsequent to 1948.  The attractive force is generally ascribed to the difference in zero-point energy of the electromagnetic field between the plates and the vacuum external to them.

**Casimir Effect**

In general, whether the Casimir force is attractive or repulsive depends on the geometry of the conducting surfaces (e.g., parallel plates or cavities of various shapes), and most importantly the boundary conditions imposed.  For example, the Dirichlet and perfectly conducting boundary conditions that are often used give very different results in a cavity having dimensions $a_1$, $a_2$ and $a_3$.  The case of perfectly conducting boundary conditions is shown in Fig. 1.

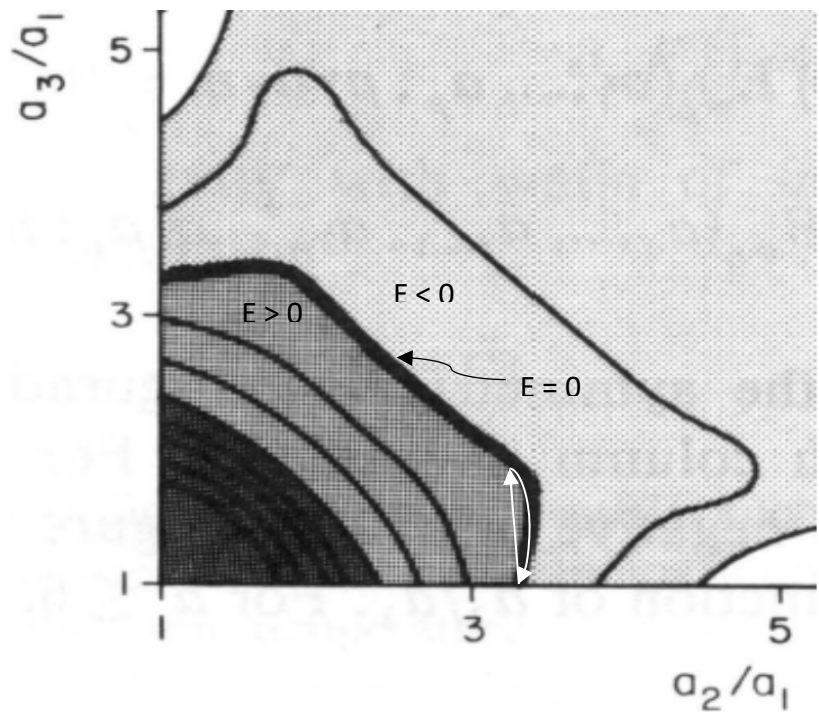

Figure 1.  Contour plot of the energy density (energy/volume) for a perfectly conducting box having dimensions $a_1$ x $a_2$ x $a_3$.  Darker shading corresponds to higher energies.  The plot gives the energy density as a function of the shape of the box. (Adapted from Fig. 4.2 of Ambjørn and Wolfram.[2])

The results of Ambjørn and Wolfram, which used dimensional regularization techniques, have generally been confirmed by Hacyan, et al.[3], although the latter note that they find that the cube is not exactly the configuration having maximum energy density and ascribe the difference as being due to Ambjørn and Wolfram's use of numerical results.

The first computation of a repulsive Casimir force was by Boyer[4] in 1968.  He computed the zero-point energy for a conducting spherical shell or radius $r$ to be $\Delta E(r) \cong 0.09\ \hbar c/2r$.  $\Delta E$ is the difference in zero-point energies for two spherical configurations.

The zero-point fluctuations have been used to explain the phenomenon of van der Waals attraction, but the existence of repulsive Casimir forces[5] would seem to invalidate this explanation.  Repulsive Casimir forces appear for the cavity configurations corresponding to the positive energy density regions shown in Fig. 1.

These positive energy density configurations are not eliminated by changing the boundary conditions of the cavity from perfectly conducting walls to the more realistic case of Dirichlet conditions as shown by Ambjørn and Wolfram in their Fig. 3.5.

Perhaps the greatest concern about the perfectly conducting wall boundary condition is that it allows for what has been called a "Casimir vacuum energy extraction cycle," shown by the cyclic path shown in white in Fig. 1.   Keeping $a_1 = 1$ constant, the cyclic path begins where $a_3 = 1$ and $a_2$ is at the intersection of $a_3 = 1$ with the E = 0 contour ($a_2 \sim 3.3$); it then goes through a series of configurations in the E > 0 region until it intersects the E = 0 contour ($a_3 \sim 1.85$) and then the configurations follow this contour to the origin of the cyclic path.  Other cyclic paths that begin on the E = 0 contour go through the E > 0 region to another point on the E = 0 contour and then return via the zero energy contour to the starting point are possible. Realistic cavities would of course require some energy expenditure to move the boundaries.  The cyclic path shown in Fig. 1 corresponds to that used by Forward as discussed in the next paragraph.

This possibility was first pointed out by R.L. Forward in the NASA Breakthrough Propulsion Physics Workshop Proceedings available from the NASA Technical Reports Server (Document ID 19990023210).  This led to several attempts to experimentally prove this possibility up until current times.  Many of these experiments were done under the auspices of the Johnson Space Center, in Houston, TX.  These are difficult experiments to perform and none —as far as I know— has given a definitive result.

**Problems with Zero-point Fluctuations and Relativistic Physics**

Often in quantum field theories the zero-point fluctuation energies are ignored or dealt with by sleight of hand.  But this is not possible in special relativity where this energy is part of a four vector.  If the vacuum state of the theory is to be invariant under Lorentz transformations the zero-point energy must vanish.

In General Relativity the problem is even worse since the energy associated with the zero-point fluctuations adds an additional term to the right-hand side of Einstein's field equations.  In the 1920s, Pauli expressed his concern about the gravitational effects of the zero-point energy, by calculating that the radius of the universe "nicht einmal bis zum Mond reichen würde" (would not even reach the Moon).

While a small cosmological constant term is needed to account for the observation that the expansion of the universe is accelerating, although this has now come into question, its magnitude is miniscule compared to the estimates of the energy density due to zero-point fluctuations.[6]

The way out of this conundrum is to reexamine whether or not the Casimir effect does prove that zero-point vacuum fluctuations are real. This was done in 2005 by Jaffe[7] and it is worth quoting the Abstract of this paper:

> "In discussions of the cosmological constant, the Casimir effect is often invoked as decisive evidence that the zero-point energies of quantum fields are "real." On the contrary, Casimir effects can be formulated and Casimir forces can be computed without reference to zero-point energies. They are relativistic, quantum forces between charges and currents. The Casimir force (per unit area) between parallel plates vanishes as $\alpha$, the fine structure constant, goes to zero, and the standard result, which appears to be independent of $\alpha$, corresponds to the $\alpha \rightarrow \infty$ limit."

Jaffe, while he does show that the Casimir effect cannot be used to prove the reality of the zero-point vacuum fluctuations, cautions that the reality of these fluctuations, should they exist, are beyond the scope of his paper and that the question of their reality remains open. However, Jaffe expresses doubt that a consistent formulation of relativistic quantum mechanics without zero-point energies exists.  Note that Jaffe did not extend his doubt to quntum field theory.

In the next couple of years after Jaffe's paper, Herzberg, Jaffe, Kardar, and Scardicchio wrote two papers[8,9] that conclude that the Casimir force is always attractive, and repulsive Casimir froces are invalidated by some "cutoff dependence".  By cutoffs the authors mean physical caracteristics of the metal used for the cavity such as plasma frequency and skin depth.  They study a metalic parallelepiped with a movable partition at some distance from the base.  They find that the finite part of the energy can only be positive (giving a repulslive force) if only one of the boxes on each side of the partition is considered, "while if both compartments are included, the net force on the partition is attractive (in the sense that it is pulled to the closest base)."  They call the configuration of the parallelepiped with the movable partition a "piston".  When one consider ideal rather than real materials, which today come with cutoffs, the authors find the results to be "unphysical and contrived".  This criticism may well be valid when it comes to past experimental attempts to extract energy from the vacuum, as discussed above, but there is no reason not to study the theoretical issues for ideal materials.

**Origin of the Zero-Point Vacuum Fluctuations**

The origin of the zero-point vacuum fluctuations is directly related to the quantization of the classical simple harmonic oscillator.  In one dimension one starts from the classical equation of motion $m\ddot{x} = -gx$ for particle with mass $m$ moving under a force -$gx$. This has the solution $x = x_0 \exp(i\omega t)$, where the frequency of oscillation is $\sqrt{g/m}$.  The Hamiltonian is then given by $H = \frac{1}{2m}p^2 + \frac{1}{2}gx^2$, where p is the momentum $m\dot{x}$.

To quantize this simple harmonic oscillator, one uses the commutation relations for the conjugate operators $x$ and $p$ given by $[x,p] = i\hbar$.  First factor the sum of squares in the Hamiltonian into the product of the operators $a$ and $a^\dagger$ defined as

$$a = \frac{1}{\sqrt{2\hbar\omega}}\left(\frac{1}{\sqrt{m}}p - i\sqrt{g}\,x\right),$$
$$a^\dagger = \frac{1}{\sqrt{2\hbar\omega}}\left(\frac{1}{\sqrt{m}}p + i\sqrt{g}\,x\right),$$

(1)

where $a^\dagger$ is the Hermitian adjoint of $a$. These operators satisfy the commutation relation $[a, a^\dagger] = 1$. The Hamiltonian may now be expressed in terms of the operators $a$ and $a^\dagger$ as $H = \frac{1}{2}\hbar\omega(aa^\dagger + a^\dagger a) = \hbar\omega\left(a^\dagger a + \frac{1}{2}\right)$, where the commutation relation for $[a, a^\dagger]$ was used. The factor of ½ in the parentheses corresponds to the zero-point energy.

Weinberg[10] has given an historical approach to quantizing a simple one-dimensional harmonic oscillator, which is said to apply to either a string whose vibrations are constrained to vanish at the origin of the coordinate $x$ and at $x = L$ or to the full electromagnetic field. Thus, the mass $m$ does not make an appearance. This approach will be used in what follows.

Consider a radiation field $u(x,t)$ in one spatial dimension. The spatial coordinate $x$ ranging from 0 to $L$. The radiation field is assumed to vanish at the endpoints 0 and $L$. The field $u(x,t)$ can then be expressed as a sum of Fourier components as

$$u(x,t) = \sum_{k=1}^\infty q_k(t)\,sin\left(\frac{\omega_k x}{c}\right),$$

where

$$\omega_k \equiv k\pi c/L.$$

(2)

The argument of the sine function might appear unusual. What has happened is that the continuous $k$-space has been changed to a lattice space and *k takes on integer values*.[11] The boundary condition has become periodic; ie., $u(x + \mathrm{L}) = u(x)$. $L$ is the "volume" of the 1-dimensional lattice space and is often set equal to unity whatever the dimension of the space being used to simplify the equations, thus obscuring the fact that one is using a lattice space. Keeping in mind that for the electromagnetic field $\lambda\nu = c$ can be written as $\omega = kc$, and then substituting the definition of $\omega_k$ into the equation for $u(x,t)$ it can be seen that $k = 1$ corresponds to one half wavelength between 0 and $L$, k = 2 to one wavelength between 0 and $L$, etc.

Recall that the Hamiltonian is the integral of the Hamiltonian density and is given in the case considered here by[12]

$$H = \frac{1}{2}\int_0^L \left\{ \left(\frac{\partial u(x,t)}{\partial t}\right)^2 + c^2\left(\frac{\partial u(x,t)}{\partial x}\right)^2 \right\} dx.$$

(3)

Substituting $u(x,t)$ from Eq. (2), yields,

$$H = \frac{L}{4}\sum_1^\infty \{\dot{q}_k{}^2(t) + \omega_k^2 q_k^2(t)\}.$$

(4)

At this point one must find a momentum conjugate to $q_k(t)$. Following Weinberg, if one considers $\dot{q}_k$ to be equivalent to the momentum, Hamilton's canonical equations tell us that the momentum is the derivative of the Hamiltonian in Eq. (4). That is,

$$\dot{q}_k(t) \equiv p_k(t) = \frac{\partial H(p_k(t), q_k(t))}{\partial p_k(t)} = \frac{\partial H(\dot{q}_k(t), q_k(t))}{\partial \dot{q}_k(t)} = \frac{L}{2}\dot{q}_k(t),$$

$$p_k(t) = \frac{L}{2}\dot{q}_k(t) \quad or \quad \dot{q}_k(t) = \frac{2}{L}p_k(t).$$

(5)

To connect this with operators in Hilbert space, use is made of the usual quantization postulate relating commutators and the classical Poisson bracket and the commutation relation between $q_k(t)$ and $p_k(t)$ becomes

$$[q_k(t), p_k(t)] = i\hbar.$$

(6)

**Explicit Solution For $q_k(t)$**

The commutator between $\dot{q}_k(t)$ and $q_k(t)$ can be obtained from the above as

$$[\dot{q}_k(t), q_k(t)] = \frac{2}{L}[p_k(t), q_k(t)] = -i\frac{2}{L}\hbar.$$

(7)

By differentiating Eq. (5) with respect to time results in

$$\ddot{q}_k(t) = \frac{2}{L}\dot{p}_k(t) = -\frac{2}{L}\frac{\partial H}{\partial q_k(t)} = -\omega_k^2 q_k(t),$$

(8)

which can be written as the differential equation

$$\ddot{q}_k(t) + \omega_k^2 q_k(t) = 0.$$

(9)

This is the equation for a simple harmonic oscillator and the formal solution is

$$q_k(t) = c_1 e^{-i\omega_k t} + c_2 e^{+i\omega_k t},$$

(10)

but there are no constraints on the arbitrary constants. To play a role in quantum field theory, they need to replaced by the operators $a_k$ and $a_k^\dagger$ that satisfy the commutation relations given by

$$[a_k, a_k^\dagger] = 1 \quad and \quad [a_k, a_j] = 0.$$

(11)

Doing so, and requiring $q_k(t)$ to satisfy the commutation relation given by Eq. (7) gives

$$q_k(t) = \sqrt{\frac{\hbar}{L\omega_k}} \left[ a_k e^{-i\omega_k t} + a_k^\dagger e^{+i\omega_k t} \right].$$

(12)

Note that $a_k$ and $a_k^\dagger$ are time-independent matrix operators. Obtaining the radical in Eq. (12) uses the form for these two operators given in Eq. (13) below. Weinberg is one of the few authors to retain the $L$ up to this point in his discussion of the birth of quantum field theory. He also gives the explicit matrices of these operators for a single normal mode and gives their physical interpretation.

To obtain the Hamiltonian in terms of the operators $a_k$ and $a_k^\dagger$ from the Hamiltonian given in Eq. (4), define $a_k$ and $a_k^\dagger$ as

$$a = \frac{1}{\sqrt{2\hbar\omega}}(\dot{q}_k(t) \pm i\omega q)$$

$$a^\dagger = \frac{1}{\sqrt{2\hbar\omega}}(\dot{q}_k(t) \mp i\omega q).$$

(13)

This is a factorization of the sum of squares in the Hamiltonian given by Eq. (4). Either the upper signs or lower signs may be used in Eq. (13) since they give the same result. Solving for $q$ and $\dot{q}$ from Eqs. (13) gives

$$q = i\sqrt{\frac{\hbar}{2\omega}} (a - a^\dagger)$$

$$\dot{q} = \sqrt{\frac{\hbar\omega}{2}} (a + a^\dagger).$$

(14)

Substituting into the expression for the Hamiltonian given in Eq. (4) results in

$$H = \sum_k \hbar\omega_k \left(a_k^\dagger a_k + \frac{1}{2}\right).$$

(15)

Note that the zero-point energy is still present.

$a_k$ and $a_k^\dagger$ are known respectively as annihilation and creation operators. The importance of expressing the Hamiltonian in terms of creation and annihilation operators is that any operator can be expressed as a sum of products of annihilation and creation operators. And if these have non-singular coefficients, the scattering or *S* matrix will satisfy what is known as the cluster decomposition principle, which states that the *S* matrix describes interactions that are at least approximately local.

To show that $a_k$ and $a_k^\dagger$ do play the role of annihilation and creation operators, define the following set of ket-vector basis functions, consistent with the commutator relation in Eq. (11), by

$$a_k \left|n> = n^{\frac{1}{2}} \right|n-1>$$
$$a_k^\dagger \left|n> = (n+1)^{\frac{1}{2}} \right|n+1>.$$

(16)

*n* designates the occupation number of a state of the system. In addition, with regard to the vacuum state,

$$a_k \left|0> = 0 \quad and. \quad <0\right|a_k^\dagger = 0.$$

(17)

The commutator maps each basis function onto itself; i.e.,

$$\left[a_k, a_k^\dagger\right]\left|n> = \right|n>.$$

(18)

Moreover, the function $|n>$ is an eigenvector of the Hamiltonian, these eigenfunctions being normal and orthogonal; i.e., $< n|n' > = \delta_{nn'}$.

All of the above is dependent on the commutator of Eq. (6), which results from relating commutators to the classical Poisson bracket. It is important to realize where the commutation relation between $q_k(t)$ and $p_k(t)$ came from.

Its origin dates back to 1925 when Heisenberg understood and explained the limitations of Bohr's theory, which depended on the quantum condition $\oint pdq = nh$. Bohr's theory required knowing the orbit of an electron around the nucleus of an atom and its velocity in that orbit. Since the position of an electron in an atom is unobservable, Heisenberg introduced his idea of matrix mechanics, later proved by Schrodinger to be equivalent to his wave mechanics.

In Heisenberg's theory, every observable physical magnitude is associated with a representative matrix and in particular the momenta $p_k$ and coordinates $q_k$. The matrices representative of these observables are not commutative and their commutator was set to be that given by Eq. (6). Originally the matrix theory was used to explain the frequency of the radiation emitted during transitions between two atomic energy levels $\nu_{nm} = (E_n/h) - (E_m/h)$. The commutator applies not only to the $p_k$ and $q_k$ associated with discrete atomic levels but also further from the nucleus in the continuum, which ultimately becomes free space; thus, the commutator is not topologically dependent on the closed energy levels near the nucleus.

**Normal Ordering**

The Hamiltonian may also be written as

$$H = \frac{1}{2}\sum_k \hbar\omega_k(a_k^\dagger a_k + a_k a_k^\dagger).$$

(19)

If normal ordering consists of putting all creation operators to the left of destruction operators, doing this in Eq. (19) gives the Hamiltonian

$$H = \sum_k \hbar\omega_k.$$

(20)

As a result, normal ordering, as stated above, eliminates the zero-point energy with its associated infinities. It is interesting to consider the commutator in Eq. (11) between $a_k$ and $a_k^\dagger$, $[a_k, a_k^\dagger] = a_k a_k^\dagger - a_k^\dagger a_k$. Simply applying the definition given above to the expanded commutator means $[a_k, a_k^\dagger] = 0$. This is what one should expect because this definition of normal ordering means that $a_k$ and $a_k^\dagger$ are treated as if they commute with each other. So, the rule is that when using this definition of normal ordering it should not be applied to commutators.

But there are in practice two definitions for normal ordering: $\mathcal{N}$, and that designated by the "double dot" operation; e.g., $:(a_k a_k^\dagger):$. In general, if $F(a, a^\dagger)$ is some *function* of the operators $a$ and $a^\dagger$, then $\mathcal{N}(F(a, a^\dagger))$ means that the commutator $[a, a^\dagger] = 1$ is used to move all creation operators to the left of destruction operators. The operator $F(a, a^\dagger)$ remains the same, it is only its functional representation that changes. The double dot operation $:F(a, a^\dagger):$ moves all creation operators to the left of destruction operators *without* taking into account of the commutator relation. As put by Blasiak,[13] the normal ordering problem for $F(a, a^\dagger)$ is solved if an operator $G(a, a^\dagger)$ can be found that satisfies

$$F(a, a^\dagger) = \mathcal{N}\left(F(a, a^\dagger)\right) \equiv\ :G(a, a^\dagger):\ .$$

(21)

In quantum field theory it is the $\mathcal{N}$ definition of normal ordering that is generally used along with the commutation or anticommutation relations to bring operators into the functional representation where all creation operators are to the left of destruction operators.

The normal ordering discussed above, the expectation values of normally ordered operators only vanish in the free theory. There is a generalized form of normal ordering called "complete normal ordering"[14] that cancels all tadpole Feynman diagrams to any order in perturbation theory. The expectation values of completely normal-ordered operators vanish identically in the full interaction theory.

**Quantum Field Theory**

Most books on quantum field theory use the approach of postulating a Lagrangian followed by using the rules of canonical quantization. However, the structure of quantum field theory is to a large extent the result of requiring quantum mechanics to satisfy special relativity as well as its associated symmetries. For this reason, as well as others, if one is not to induce cognitive dissonance into the minds of those learning quantum field theory it is best to reject this approach and, as was suggested by Weinberg in his book on the subject, use one that is ahistorical. In the end, one can find that most free field theories provide operators $q^n(\boldsymbol{x},t)$ and their canonical conjugates $p_n(\boldsymbol{x},t)$ that satisfy the usual canonical commutation or anticommutation relations.

<div align="center">

**Summary**

</div>

Most texts on quantum mechanics and quantum field theory express some angst at eliminating the vacuum energy by normal ordering or informal hand waving, or, at best as stated by Schweber, Bethe, and Hoffman, eliminating the zero-point energy "can actually be justified, by the fact that in the transition from classical to quantum theory the order of the operators is not defined to terms of order $\hbar$" (they reference a 1931 paper by Rosenfeld and Solomon). But, thanks to Jaffe, there is no longer any reason for trying to justify its elimination. There is no proof whatsoever that zero-point oscillations or, in quantum field theory, that Feynman diagrams having loops with only a single vertex (so called tadpole graphs) exist. And in the latter case, they can be eliminated by complete normal ordering to any order in perturbation theory.

The implications of this for cosmology in particular are dramatic. As pointed out by Weinberg, the "cosmological constant problem" is that its observational limits differ from theoretical expectations by some 120 orders of magnitude.  In the thirty years since he wrote his paper referenced above, not much has changed.  But the problem becomes much more tractable if there are no contributions from zero-point energies.  While observations of type 1a supernovae in 1998 appeared to show there was an accelerating expansion of the universe, which would still imply there is some cosmological "problem", recent work using observations of many more supernovae concluded there was little evidence for isotropic acceleration.[15]  As put by Colin, et al., "the cosmic acceleration deduced from supernovae may be an artefact of our being non-Copernican observers, rather than evidence for a dominant component of 'dark energy' in the Universe".  Their results are consistent with no acceleration.  Thus, there appears to be no cosmological constant problem.

Perhaps the most enigmatic cosmological problem left is that which comes from the rotation curves of galaxies indicating the existence of collisionless cold dark matter of an unknown nature. Such matter also dominates the mass content of galaxy clusters.  The density profiles of galaxy dark matter halos are often modeled by an approximate solution to the isothermal Lane-Emden equation with appropriate boundary conditions at the origin.  It has been shown that such a model corresponds to an exact solution of the Einstein-Maxwell equations for charged dust.[16] This could indicate that there might exist particles having the unusual nature of having an exotic charge of only one sign which interacts only with other such charged particles, and gravitationally with normal matter and other particles like itself.